\documentclass[prc,twocolumn,superscriptaddress, showpacs,amsmath,amssymb,floatfix,nofootinbib]{revtex4-1}
\usepackage{graphicx,bm,url,braket,xcolor,xspace}
\usepackage{hyperref}
\usepackage[normalem]{ulem}
\newcommand{\bb}{$\beta\beta$\xspace}
\newcommand{\bbz}{$0\nu\beta\beta$\xspace}

\begin{document}

\title{Proton-Neutron Pairing Amplitude as a Generator Coordinate for
Double-Beta Decay}

\author{Nobuo Hinohara}
\affiliation{Department of Physics and Astronomy, University of North Carolina,
Chapel Hill, North Carolina, 27599-3255, USA}
\affiliation{Center for Computational Sciences, University of Tsukuba, Tsukuba
305-8571, Japan}

\author{Jonathan Engel}
\affiliation{Department of Physics and Astronomy, University of North Carolina,
Chapel Hill, North Carolina, 27599-3255, USA} 

\begin{abstract}
We treat proton-neutron pairing amplitudes, in addition to the nuclear
deformation, as generator coordinates in a calculation of the neutrinoless
double-beta decay of $^{76}$Ge.  We work in two oscillator shells, with a
Hamiltonian that includes separable terms in the quadrupole, spin-isospin, and
pairing (isovector and isoscalar) channels.  Our approach allows larger
single-particle spaces than the shell model and includes the important physics
of the proton-neutron quasiparticle random-phase approximation (QRPA) without
instabilities near phase transitions.  After comparing the results of a
simplified calculation that neglects deformation with those of the QRPA, we
present a more realistic calculation with both deformation and proton-neutron
pairing amplitudes as generator coordinates.  The future should see
proton-neutron coordinates used together with energy-density functionals.
\end{abstract}
\maketitle

Neutrinoless double-beta (\bbz) decay can occur only if neutrinos are Majorana
particles.  That fact has motivated many expensive and complicated experiments
to search for the process.  If it is observed, the decay can also reveal an
overall neutrino mass scale, $m_\nu \equiv \sum_i U_{ei} m_i^2$ (where $i$
labels the mass eigenstates, and $U$ is the neutrino mixing
matrix~\cite{avi08}), but only if we know the value of unobservable nuclear
matrix elements that play a role in the decay. The entanglement of nuclear and
neutrino physics has led nuclear-structure theorists to attempt to calculate
the nuclear matrix elements.  At present, various nuclear models agree to
within factors of about three.  More accurate calculations will increase the
importance of both existing limits and any actual observation.  

The method most often applied to double-beta decay is the proton-neutron
(\textit{pn}) quasiparticle random phase approximation (QRPA). QRPA
calculations were the first to suggest that \textit{pn} pairing quenches
double-beta matrix elements \cite{PhysRevLett.57.3148,PhysRevC.37.731}.  That
result was surprising because evidence for such pairing in spectra and
electromagnetic transitions or moments is hard to come by, particularly when
the number of neutrons $N$ is significantly larger than the number of protons
$Z$, as it is in most nuclei that undergo double-beta decay.  At the mean-field
level, \textit{pn} pairing never develops in those nuclei. But the QRPA
uncovered pairing fluctuations that have significant effects on both single-
and double-beta decay. 

Despite this success, the QRPA is not fully realistic because it is built on
small oscillations around a single mean field.  That means that 1) it is not
really suited for complicated but important double-beta-decay parents/daughters
such as $^{76}$Ge and $^{130}$Xe, in which a single mean field provides a poor
representation of the ground state, and 2) its predictions for the effects of
\textit{pn} pairing fluctuations, which are not small, cannot be fully
trusted.  To treat the physics more reliably, one needs a method in which
collective \textit{pn} pairing fluctuations are allowed to be large.  One
framework for large-amplitude modes is the generator-coordinate method (GCM)
\cite{rin04,RevModPhys.75.121}, a variational procedure that works by mixing
many mean-field wave functions with varying amounts of collectivity.  To treat
large-amplitude quadrupole vibrations, for example, it produces a ``collective
wave function'' that superposes Slater determinants (or generalizations) with
different amounts of quadrupole deformation in an optimal way.  In our work the
\textit{pn} pairing amplitude (defined below) will play the role of collective
deformation.

In fact, a sophisticated version of the GCM, in conjunction with energy-density
functional theory, has already been applied to double-beta decay
\cite{rod10,rod11,vaq13}.  The collective coordinates include only axial
quadrupole deformation and particle-number fluctuation (from like-particle
pairing), however, and so the calculations omit the suppression caused by
\textit{pn} pairing.  Not surprisingly, the GCM results tend to be larger than
those of e.g.\ the shell model, which includes \textit{pn} pairing.  If the
\textit{pn} pairing amplitude could be added as another coordinate in these GCM
calculations, the matrix elements would probably shrink.  No one has ever
treated \textit{pn} pairing as a GCM degree of freedom, however.  Because
\textit{pn} pairing is less strongly collective than its like-particle
counterpart, doing so requires a careful extension of mean-field methods and
the GCM itself.  In this paper we undertake that project and report a first
application to the \bbz decay of $^{76}$Ge.

We begin with the matrix elements we wish to calculate.  In the closure
approximation (which is good to about 10\% \cite{pan90}), we can write the
nuclear \bbz matrix element in terms of the initial and final ground states and
a two-body transition operator. If we neglect the ``tensor term,'' the effect
of which is at most 10\% \cite{kor07a,men08}, and two-nucleon currents
\cite{men11} (the effects of which are still uncertain) we can write the matrix
element as
\begin{align} 
\label{eq:me}
M^{0\nu} &\equiv \bra{F}\hat{M}_{0\nu}\ket{I} =  \frac{2R}{\pi g_A^2} \int_0^\infty \!\!\! q \, dq  \\ 
&\hspace{-1cm}\times \bra{F} \sum_{a,b}\frac{j_0(qr_{ab})\left[ h_{\rm F}(q)+   
 h_{\rm GT}(q) \vec{\sigma}_a \cdot \vec{\sigma}_b \right]}
 {q+\overline{E}-(E_I+E_F)/2} \tau^+_a \tau^+_b \nonumber \ket{I} \,,
\end{align}
where $\ket{I}$ and $\ket{F}$ are the ground states of the initial and final
nuclei, $r_{ab}$ is the distance between nucleons $a$ and $b$, $j_0$ is the
usual spherical Bessel function, and $R$ is the nuclear radius, inserted by
convention to make the matrix element dimensionless.  The form factors $h_{\rm
F}(q)$ and $h_{\rm GT}(q)$ contain the vector and axial vector coupling
constants, forbidden corrections to the weak current, nucleon form factors, and
the ``Argonne'' short-range correlation function \cite{sim09}.  See, e.g.,
Ref.\ \cite{sim08} for details; note that we absorb the inverse square of the
axial-vector coupling constant into our definition of $h_F$.

To compute the matrix element in Eq.\ (\ref{eq:me}) we need good
representations of the initial and final ground states $\ket{I}$ and
$\ket{F}$.  In this first application to $A=76$ nuclei, we construct the states
in a Hilbert space consisting of 36 particles moving freely in the oscillator
$fp$ and $sdg$ shells.  Our Hamiltonian has the form 
\begin{align}
\label{eq:h}
H &=h_0 -\sum_{\mu=-1}^1  g^{T=1}_\mu \ S^\dag_\mu S_\mu -\frac{\chi}{2}
\sum_{K=-2}^2 Q^\dag_{2K} Q_{2K} \nonumber \\ 
&- g^{T=0} \sum_{\nu=-1}^1 P^\dag_\nu P_\nu + g_{ph} \sum_{\mu,\nu=-1}^1
F^{\mu\dag}_\nu F^\mu_\nu \,,
\end{align}
where $h_0$ contains spherical single particle energies, $Q_{2K}$ are the
components of a quadrupole operator defined in Ref.\ \cite{Baranger1968490}, and
\begin{eqnarray}
\label{eq:h-ops}
S^\dag_\mu  &=& \frac{1}{\sqrt{2}} \sum_l \hat{l} [c^\dag_l
c^\dag_l]^{001}_{00\mu} \,, \quad P^\dag_\mu = \frac{1}{\sqrt{2}} \sum_l 
\hat{l} [c^\dag_l c^\dag_l]^{010}_{0\mu 0} \,, \nonumber \\ 
F^{\mu}_{\nu} &=& \frac{1}{2} \sum_{i}\sigma^{\mu}_i
\tau^{\nu}_i
= \sum_{l}\hat{l}[c^\dag_l \bar{c}_l]^{011}_{0\mu\nu} \,. 
\end{eqnarray}
In this last equation, $c^\dag_l$ is a creation operator, $l$ labels
single-particle multiplets with good orbital angular momentum, $\hat{l} \equiv
\sqrt{2l+1}$, $S^\dag_\mu$ creates a correlated isovector pair with orbital
angular momentum $L=0$ and spin $S=0$ (and with $\mu$ labeling the isospin
component $T_z$), $P^\dag_\mu$ creates an isoscalar \textit{pn} pair with $L=0$
and $S=1$ ($S_z=\mu$), and the $F^\mu_\nu$ are the components of the
Gamow-Teller operator. Although the Hamiltonian is not fully realistic, it
combines and extends both the $SO(8)$ model \cite{eng97,Dussel1986164} and the
pairing-plus-quadrupole model \cite{Baranger1968490,RevModPhys.35.853}, and
contains the most important (collective) parts of shell-model interaction
\cite{duf96}.  We discuss the values of the couplings in Eq.\ (\ref{eq:h})
shortly. 

A direct diagonalization in a space this large is not possible, even with our
simple Hamiltonian, and we have already discussed the drawbacks of the QRPA.
We therefore turn to the GCM, which has been reviewed in many places (see,
e.g., Ref.\ \cite{rin04}) and is useful in very-large-scale shell-model
problems.  The procedure is variational, with an ansatz for the ground state of
the form
\begin{equation}
\label{eq:gcm-ansatz}
\ket{\Psi}= \sum_{a_1a_2\ldots a_n} f(a_1,a_2,\ldots,a_n) \mathcal{P} \ket{a_1,a_2,\ldots,a_n} \,.
\end{equation}
Here the kets $\ket{a_1,a_2,\ldots,a_n}$ are mean-field states --- Slater
determinants or, in our case, quasiparticle vacua --- with $n$ expectation
values $a_i$ specified, $\mathcal{P}$ is an operator that projects onto states
with well-defined values for angular momentum and neutron and proton particle
numbers, and $f$ is a weight function.  The starting point, if we want to
include the effects of \textit{pn} pairing, is a Hartree-Fock-Bogoliubov (HFB)
code that mixes neutrons and protons in the quasiparticles, i.e.\
(schematically): 
\begin{equation}
\label{eq:pnhfb}
\alpha^\dag \sim u_p c^\dag_p + v_p c_p + u_n c^\dag_n + v_n c_n \,.
\end{equation}
The actual equations contain sums over single particle states as well, so that
each of the $u$'s and $v$'s above are replaced by matrices as described, e.g.,
in Ref.\ \cite{goo79}.

We use the generalized HFB (neglecting the Fock terms in this step) without any
symmetry restriction to construct a set of quasiparticle vacua that are
constrained to have a particular deformation $\beta$ (defined here as 0.438
fm$^2$ MeV$^{-1}$ $\chi \braket{Q_{20}}$) and isoscalar-pairing amplitude
$\phi=\braket{P_0+P_0^\dag}/2$ (these are the $a_i$ in Eq.\
(\ref{eq:gcm-ansatz})), that is, we solve the HFB equations for the Hamiltonian
with linear constraints
\begin{equation}
\label{eq:constrained-hfb}
H'=H - \lambda_Z N_Z - \lambda_N N_N - \lambda_Q Q_{20} -
\frac{\lambda_P}{2} \left( P_0 + P_0^\dag \right) \,,
\end{equation}
where the $N_Z$ and $N_N$ are the proton and neutron number operators --- they
are part of the usual HFB minimization --- and the other $\lambda$'s are
Lagrange multipliers to fix the deformation and isoscalar pairing amplitude.
(When computing the Fermi part of the \bbz matrix element we substitute the
isovector \textit{pn} operators $(S_0-S^\dag_0)/2i$ for $(P_0+P_0^\dag)/2$ in
Eq.\ (\ref{eq:constrained-hfb}).) As already noted, without the last multiplier
the isoscalar pairing amplitude vanishes unless the strength $g^{T=0}$ of the
corresponding interaction is larger than some critical value.  For realistic
Hamiltonians that is never the case, hence the need to generate amplitudes by
force, as it were.

Having obtained a set of HFB vacua with varying amounts of axially symmetric
deformation and \textit{pn} pairing, we project the vacua onto states with the
correct number of neutrons and protons and with angular momentum zero.  We then
solve the Hill-Wheeler equation \cite{rin04}, which amounts to diagonalizing
$H$ in the space spanned by our nonorthogonal projected vacua, to determine the
weight function $f$ in Eq.\ (\ref{eq:gcm-ansatz}).

\begin{table}[t]
\caption{ \label{tab:spenergy} Neutron and proton canonical single-particle
energies (in MeV) taken from spherical HFB with SkO$'$.}
\begin{ruledtabular}
\begin{tabular}{ccccc}
     & $^{76}$Ge (n) & $^{76}$Ge (p)& $^{76}$Se (n) & $^{76}$Se (p)\\ \hline
p1/2 & -10.31 & -6.80 & -11.21 & -5.06 \\
p3/2 & -12.69 & -9.56 & -13.72 & -7.81 \\
f5/2 &  -9.94 & -7.47 & -11.08 & -5.61 \\
f7/2 & -17.63 & -15.90& -18.87 & -13.95\\ \hline
s1/2 &  -0.49 &  6.27 &  -0.90 &  7.05  \\
d3/2 &   1.49 &  7.94 &   1.17 &  8.77  \\
d5/2 &  -2.60 &  2.64 &  -3.26 &  3.65  \\
g7/2 &   3.23 &  6.43 &   2.24 &  8.08  \\
g9/2 &  -8.09 & -6.31 &  -9.31 & -4.46  \\   
\end{tabular}
\end{ruledtabular}
\end{table}

To carry out a fairly realistic calculation, we need appropriate values for the
couplings in the Hamiltonian of Eq.\ (\ref{eq:h}).  We determine them by trying
to reproduce the results of calculations with two different Skyrme interactions
(SkO$'$ \cite{PhysRevC.60.014316} and SkM* \cite{Bar82}) in $^{76}$Ge and
neighboring nuclei.  We first do Skyrme-HFB calculations
\cite{Stoitsov20131592} in $^{76}$Ge to determine appropriate volume pairing
constants.  We then take single-particle energies for each nucleus, which we
show for SkO$'$ in Table \ref{tab:spenergy}, from the results of constrained
HFB calculations for $^{76}$Ge and $^{76}$Se, which we temporarily force to be
spherical.  Next we adjust the like-particle part of our isovector pairing
interaction ($g_1^{T=1}$ and $g_{-1}^{T=1}$) to get the same pairing gaps as
the original Skyrme calculations. The resulting occupation numbers are close to
the spherical Skyrme-HFB numbers (and, as is typical for such calculations,
relatively different from the measured occupations of Refs.\
\cite{sch08,sch09}).  The Coulomb interaction is not included explicitly in our
Hamiltonian, but its effects are present in single-particle energies and the
fit isovector pairing interaction.  

\begin{figure}[t]
\centering
\includegraphics[width=\columnwidth]{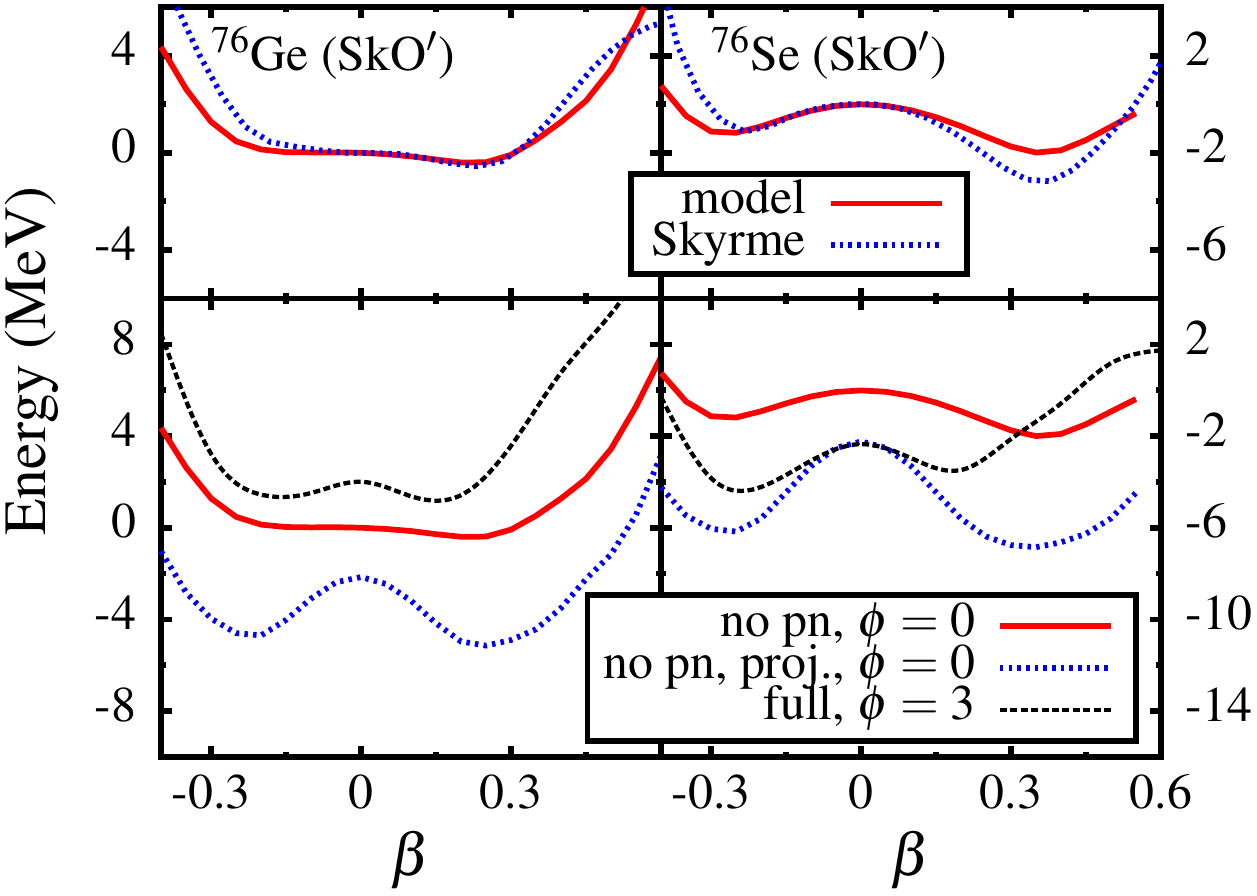}
\caption{ (Color online.) \textbf{Upper panels:} Potential energy surfaces for
SkO$'$ (dotted line) and our corresponding model interaction (solid line) at
$\phi=0$ with no spin-isospin or isoscalar pairing interactions (together
denoted by ``\textit{pn}'' in the lower panels) as functions of $\beta$ for
$^{76}$Ge (left) and $^{76}$Se (right).  \textbf{Lower panels:} Potential
energy surface from the model without projection and \textit{pn} at $\phi=0$
(solid red line, same line as in upper figure), with projection but still
without \textit{pn} at $\phi=0$ (dotted blue line), and with both projection
and \textit{pn} at $\phi=3$ (dashed black line).  Energy is measured from the
unprojected value at $\beta=0$.  \label{fig:PES} }
\end{figure}

Next, we fix our quadrupole interaction so that it reproduces the prolate
deformation of Skyrme-HFB calculations, now with axial deformation allowed, in
$^{76}$Ge and $^{76}$Se.  The top panel of Fig.\ \ref{fig:PES} compares the
diagonal part of the Hamiltonian kernel $\bra{\beta,\phi=0}H\ket{\beta,\phi=0}$
as a function of $\beta$ (we refer this function as a potential energy surface)
in the Skyrme HFB and in our model.  The lowest minima are prolate in both
nuclei.  In $^{76}$Se the surfaces have oblate minima around $\beta=-0.2$ as
well, but in general the surfaces are quite flat.  (We discuss the bottom part
of the figure later.)

Turning to the particle-hole spin-isospin interaction, we take $g_{ph}$ from a
deformed Skyrme-QRPA calculation \cite{Mustonen:2014bya}, with the relevant
piece of the time-odd functional adjusted as in Ref.\ \cite{ben02} to put the
Gamow-Teller resonance in $^{76}$Ge at the correct energy.  The resulting
values of $g_{ph}$, extracted as in Ref.\ \cite{sar98}, are $1.9 \times
\bar{g}^{T=1}$ for SkO$'$ and $0.9 \times \bar{g}^{T=1}$ for SkM*, where
$\bar{g}^{T=1}$ is the average of the two like-particle pairing strengths.  To
fix the \textit{pn} part of our $T=1$ pairing interaction, we adjust
$g_0^{T=1}$ to make two-neutrino Fermi decay vanish (in the closure
approximation); this last step approximates isospin restoration \cite{sim13}.
We find $g_0^{T=1} =$ $1.05 \times \bar{g}^{T=1}$ for SkO$'$ and $0.98 \times
\bar{g}^{T=1}$, for SkM*.

That then leaves just the crucial isoscalar pairing strength, $g^{T=0}$.  There
is no consensus about how best to determine that parameter.  We do so by
fitting the measured total $\beta^+$ strength $B({\rm GT}+)$ from $^{76}$Se as
well as possible.  Two separate charge-exchange experiments
\cite{PhysRevC.55.2802,PhysRevC.78.044301} have tried to extract $B({\rm
GT}+)$.  Neither isolates the quantity perfectly but both are consistent with
the assumption that $B({\rm GT}+) \approx 1$, and we adopt that value.  It is
not obvious how much the experimental strength is quenched by states outside
the model space, so we always do two fits, one (unquenched) as just described
and one (quenched) in which we scale our calculated strength by $(1/1.27)^2 =
0.62$.

Of course, the value we extract for $g^{T=0}$ will depend on our choice of
generator coordinates as well as assumptions about quenching.  Before turning
to the full calculation outlined above we discuss a simpler and more
transparent case, in which the quadrupole force is turned off and the isoscalar
pairing amplitude (or isovector \textit{pn} pairing amplitude when we calculate
the Fermi matrix element) is the sole coordinate.  Though the isoscalar pairs
create a spin vector that always breaks rotational symmetry and forces us to do
angular-momentum projection, the absence of a quadrupole force makes both the
initial and final nuclear densities nearly spherical.  The main advantage of
spatial spherical symmetry is that we can compare the results with those of the
spherical QRPA, for which we developed a code.  

With a single generator coordinate and the interaction we extract from SkO$'$
(minus the quadrupole part), there is no value of $g^{T=0}$ for which $B({\rm
GT}+)$ from $^{76}$Se is as small as 1, much less 0.62; we therefore let
$g^{T=0}=1.47 \times \bar{g}^{T=1}$, the value for which $B({\rm GT}+)$ is the
smallest.  With the interaction we extract from SkM*, whether we quench our
strength or not, there are two values of $g^{T=0}$ that produce the correct
$\beta^+$ strength --- 0.82 and 1.56 (unquenched) or 0.33 and 1.77 (quenched),
all in units of $\bar{g}^{T=1}$ --- and we choose the larger value in each
case.  With all parameters finally determined, we can calculate the \bbz matrix
element; Table \ref{tab:vals} displays the results at various stages of
approximation.  

In our QRPA calculation, we adjust $g^{T=0}$ (commonly called $g_{pp}$ when
divided by $\bar{g}^{T=1}$) in exactly the same way.  The values we obtain are
only slightly different.  The last column of Table \ref{tab:vals} contains the
QRPA \bbz matrix elements.  They are fairly close to those of the GCM
calculation, but much more sensitive to $g^{T=0}$. 

\begin{table}[t] 
\centering
\caption{\label{tab:vals} The \bbz matrix element $M^{0\nu}$ for the decay of
$^{76}$Ge in a simplified calculation that neglects deformation, at various
levels of approximation.  The first column contains the source of the couplings
in Eq.\ (\ref{eq:h}), the second the matrix element when the spin-isospin and
isoscalar pairing interactions are absent, the third the matrix element with
only isoscalar pairing missing, the fourth the full GCM result, and the last
the result of the QRPA with the same Hamiltonian (except for a slightly
modified $g^{T=0}$).  The matrix elements in parentheses are obtained by
quenching our $B({\rm GT}+)$.}
\begin{ruledtabular}
\begin{tabular}{lcccc}
Skyrme &  no $g_{ph}, g^{T=0}$ & no $g^{T=0}$  & full & QRPA \\[1ex]
\hline
SkO$'$    & 14.0 & 9.5 & 5.4 (5.4) & 5.6 (5.0) \\
SkM* & 11.8 & 9.4 & 4.1 (2.8) & 3.5 (2.5)\\
\end{tabular}
\end{ruledtabular}
\end{table}

\begin{figure}[t]
\centering
\includegraphics[width=\columnwidth]{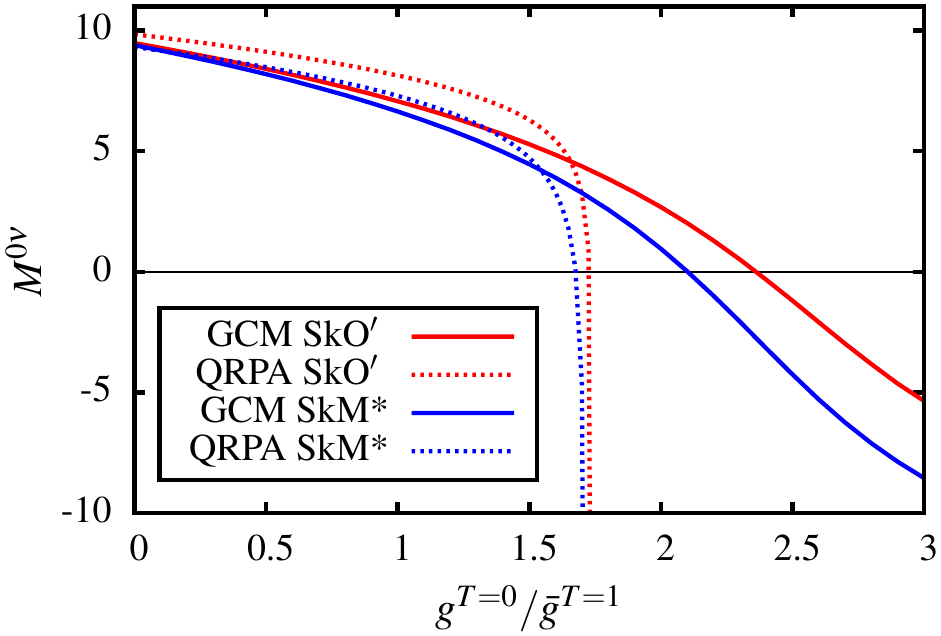}
\caption{(Color online.) Dependence of the GCM (solid) and QRPA (dashed) \bbz
matrix elements on the strength $g^{T=0}$ of the isoscalar pairing
interaction.  The red (upper) and blue (lower) lines of each type correspond to
the interaction parameters extracted from SkO$'$ and SkM*.  The divergence in
the QRPA near $g^{T=0}/\bar{g}^{T=1}=1.5$ is discussed in
the text. \label{fig:1dcurves} }
\end{figure}

To clarify this last statement, we show the GCM and QRPA matrix elements as
functions of $g^{T=0}/\bar{g}^{T=1}$ in Fig.\ \ref{fig:1dcurves}.  The QRPA
curves lie slightly above their GCM counterparts until $g^{T=0}/\bar{g}^{T=1}$
reaches a critical value slightly larger than 1.5; at that point a mean-field
phase transition from an isovector pair condensate to an isoscalar condensate
causes the famous QRPA ``collapse.'' The collapse is spurious, as the GCM
results show.  Its presence in mean-field theory makes the QRPA unreliable near
the critical point.  It is actually a bit of a coincidence that the QRPA matrix
elements in the table are as close as they are to those of the GCM; a small
change in $g^{T=0}$ would affect them substantially (though because it also
alters $B({\rm GT}+)$ a lot, fitting to $B({\rm GT}+) = 0.62$ rather than 1.0
does not have a huge effect on the \bbz matrix element).  The GCM result is not
only better behaved near the critical point but also, we believe, quite
accurate.  In the $SO(8)$ model used to test many-body methods in \bb decay
many times, the GCM result is nearly exact for all $g^{T=0}$.  That is not the
case for extensions of the QRPA that attempt to ameliorate its shortcomings
\cite{toi95,PhysRevC.66.051303}, though some of those work better around the
phase transition than others.  

\begin{figure}[b]
\centering
\includegraphics[width=\columnwidth]{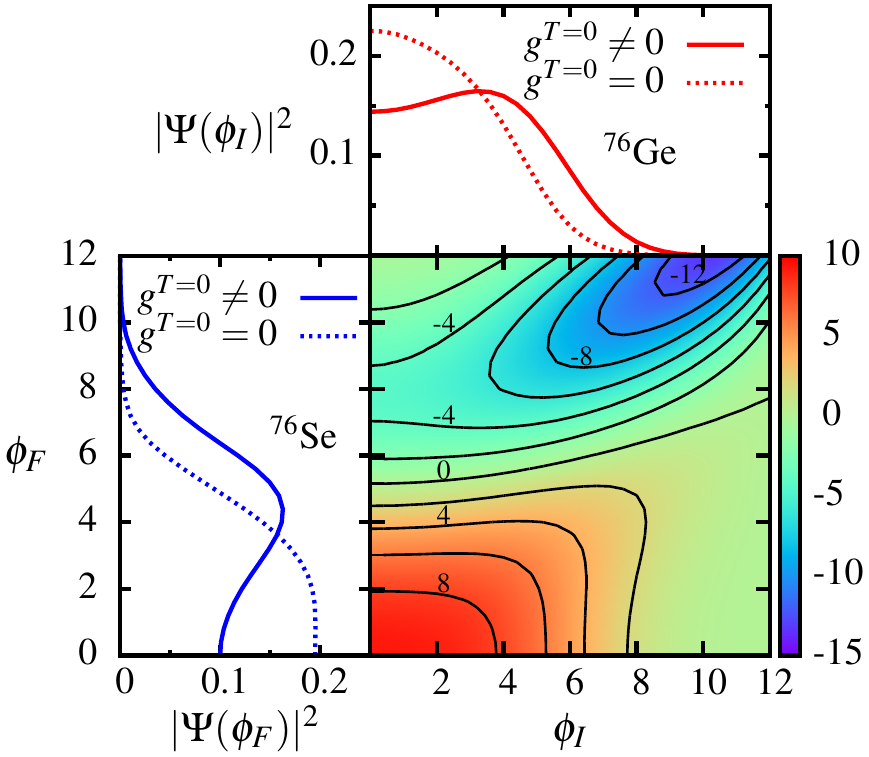}
\caption{(Color online.) \textbf{Bottom right:} $\mathcal{N}_{\phi_I}
\mathcal{N}_{\phi_F}\bra{\phi_F} \mathcal{P}_F \hat{M}_{0\nu} \mathcal{P}_I
\ket{\phi_I}$ for projected quasiparticle vacua with different values of the
initial and final isoscalar pairing amplitudes $\phi_I$ and $\phi_F$, from the
SkO$'$-based interaction (see text).  \textbf{Top and bottom left:} Square of collective
wave functions in $^{76}$Ge and $^{76}$Se.
\label{fig:contour} }
\end{figure}

To show why the GCM behaves well, we display in the bottom right part of Fig.\
\ref{fig:contour} the quantity $\mathcal{N}_{\phi_I}
\mathcal{N}_{\phi_F}\bra{\phi_F}\mathcal{P}_F \hat{M}_{0\nu} \mathcal{P}_I
\ket{\phi_I}$, where $\ket{\phi_I}$ is a quasiparticle vacuum in $^{76}$Ge
constrained to have isoscalar pairing amplitude $\phi_I$, $\phi_F$ is an
analogous state in $^{76}$Se, $\mathcal{P}_I$, $\mathcal{P}_F$ project onto
states with angular momentum zero and the appropriate values of $Z$ and $N$,
and $\mathcal{N}_{\phi_I}, \mathcal{N}_{\phi_F}$ normalize the projected
states.  This quantity is the contribution to the \bbz matrix element from
states with particular values of the initial and final isoscalar pairing
amplitudes.  The contribution is positive around zero condensation in the two
nuclei and negative when the final pairing amplitude is large.  Thus the GCM
states must contain components with significant \textit{pn} pairing when
$g^{T=0}$ is near its fit value.  The appearance of this plot is different from
those in which the matrix element is plotted versus initial and final
deformation \cite{rod10,rod11,vaq13}.  Here the matrix element is small or
negative even if the initial and final pairing amplitudes have the same value,
as long as that value is large.  The behavior reflects the qualitatively
different effects of isovector and isoscalar pairs on the matrix element
\cite{PhysRevC.37.731}, effects that have no analog in the realm of deformation.

The weight function $f$ in the GCM ansatz multiplies non-orthogonal states and
so is not really a ``collective ground-state wave function.'' The object that
does play that role for is a member of an orthogonalized set defined, e.g., in
Refs.\ \cite{rin04} and \cite{rod11}.  The top and left parts of Fig.\
\ref{fig:contour} show the square of this collective wave function for
$^{76}$Ge and $^{76}$Se, with $g^{T=0}$ set both to zero and the fit value.  It
is clear in both nuclei, but particularly in $^{76}$Se, that the isoscalar
pairing interaction pushes the wave function into regions of large $\phi$,
where the matrix element in the bottom right panel is significantly reduced.
It is also clear that for $g^{T=0} \ne 0$ the collective wave functions are far
from the Gaussians that one would obtain in the harmonic (QRPA) approximation.
Isoscalar pairing really is, and must be treated as, a large-amplitude mode. 

We turn finally to the more realistic calculation that includes both
deformation and the \textit{pn} pairing amplitude as generator coordinates.  We
fit the couplings in $H$ just as described earlier; the strength of the
quadrupole interaction no longer vanishes and some of the other parameters
change slightly: $g_0^{T=1}= 0.90 $ for the interaction based on SkO$'$ and
0.79 for that based on SkM*, and $g^{T=0} = 1.75$ for SkO$'$ and 1.51 for SkM*,
in units of $\bar{g}^{T=1}$.  The calculated $B({\rm GT}+)$ in both cases is
larger than the experimental data with or without quenching, which therefore
does not alter the value of $g^{T=0}$.

First we analyze the influence of the number and angular-momentum projection on
energy.  The bottom part of Fig.\ \ref{fig:PES} shows the projected potential
energy surfaces $\bra{\beta,\phi}\mathcal{P}H\mathcal{P}\ket{\beta,\phi}$ for
two values of $\phi$, along with the unprojected surface from the top part of
the panel.  Projecting at $\phi=0$ without including \textit{pn} interactions,
the figure shows, lowers the energy by several MeV.  The correlation energy
from the angular momentum projection is large in the deformed regions, and
projection causes both the oblate and prolate configurations that are low lying
before projection to become clear minima. 

With the \textit{pn} interactions included, we present the surface at $\phi=3$,
where the collective wave function peaks, rather than at $\phi=0$.  The curve
is shifted up by the repulsive spin-isospin interaction and downward by the
isoscalar pairing, so that the final location depends on the relative sizes of
$g_{ph}$ and $g^{T=0}$.  The SkO$'$-based interaction has a particularly large
$g_{ph}$ and so the final curve is higher than the initial unprojected curve
without the \textit{pn} terms.  The curves flatten and in $^{76}$Se the oblate
minimum becomes the lowest.

\begin{table}
\caption{GCM ground state energies, in MeV, with both deformation and isoscalar
pairing as generator coordinates.  The energies are measured from the energy of
the state with $\beta=0$, $\phi=0$, $g_{ph}=g^{T=0} =0$ and no projection.  
\label{tab:GCMenergy}}
\begin{ruledtabular}
\begin{tabular}{ccccc}
 & SkO$'$ &  & SkM* \\ 
 & $^{76}$Ge & $^{76}$Se & $^{76}$Ge & $^{76}$Se \\ \hline
no $g_{ph},g^{T=0}$ & -6.0 & -8.1 & -5.5 & -7.1 \\
no $g^{T=0}$       & +10.5& +6.6 & +2.1 & -0.9 \\
full & -0.9 & -6.9 & -7.7 & -12.4
\end{tabular}
\end{ruledtabular}
\end{table}

Table~\ref{tab:GCMenergy} shows the GCM correlation energies themselves with
successively more \textit{pn} interaction included.  The spin-isospin term in
the Hamiltonian increases the energies by about 15 MeV for the SkO$'$-based
interaction, which again, is quite strong in that channel, and 8 MeV for the
SkM*-based interaction. The isoscalar pairing interaction then decreases the
energy by 10--13 MeV, depending on the nucleus and interaction.


\begin{figure}[b]
\centering
\includegraphics[width=\columnwidth]{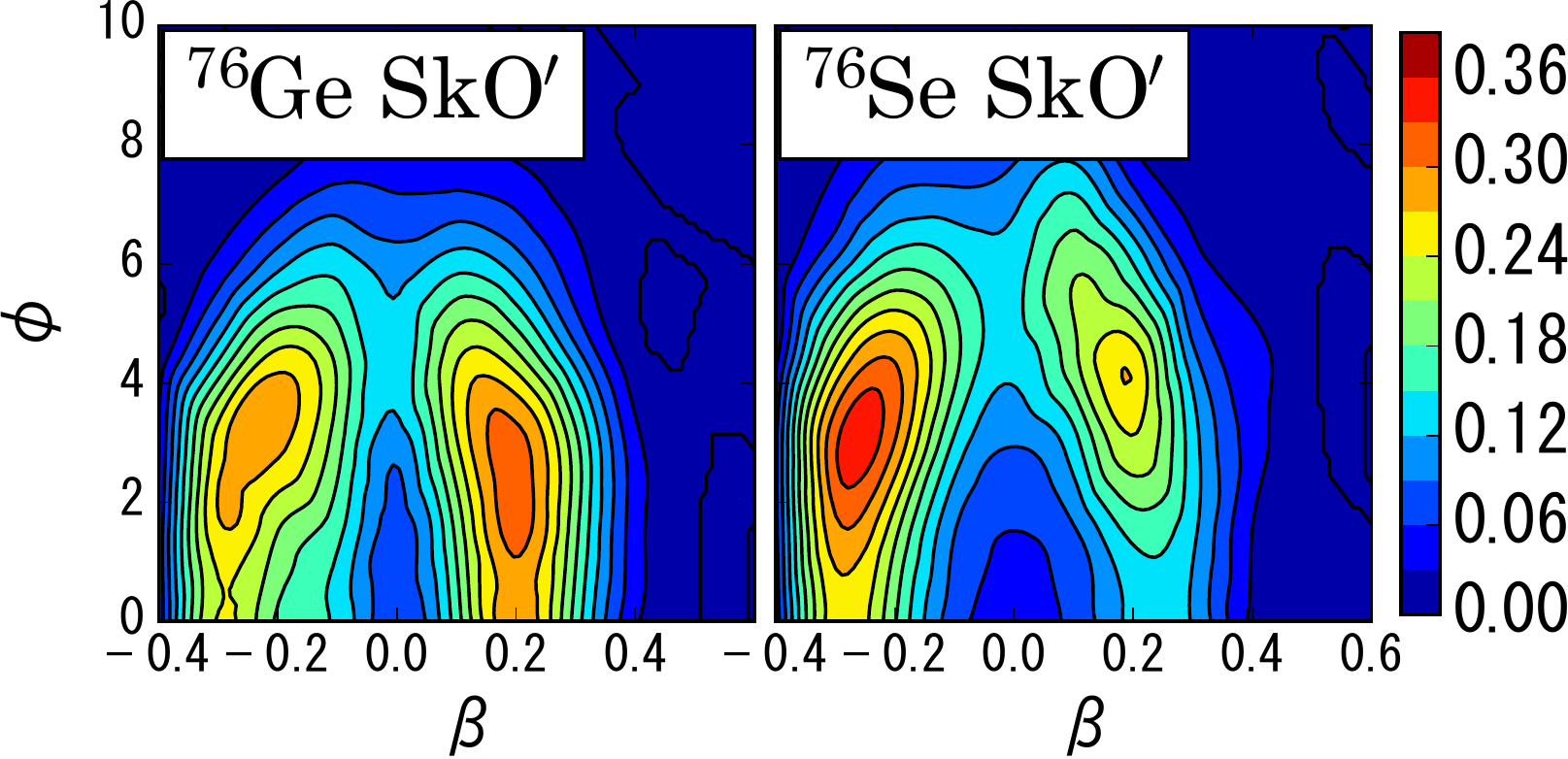}  
\caption{(Color online.) Square of the collective wave functions in the
calculation that includes deformation, in $^{76}$Ge (left) and $^{76}$Se
(right), for the SkO$'$-based interaction.
\label{fig:2dwave}
}  
\end{figure}

Finally, Fig.\ \ref{fig:2dwave} shows the squares of the collective wave
functions in $\beta$ and $\phi$.  These wave functions closely mirror the
projected potential energy surfaces.  As in the example without deformation,
the peaks are at nonzero isoscalar-pairing amplitude.  Regarding deformation,
the largest peak is in the prolate region for $^{76}$Ge and in the oblate
region for $^{76}$Se.  Though that is also the case in the calculations of
Refs.~\cite{rod10,rod11,vaq13}, our wave functions are still quite different
from the ones in those papers, and our matrix element is less suppressed.  The
full result of our calculation is $M^{0\nu}= 4.7$, with both the SkO$'$- and
SkM*-based interactions.  The number breaks down into 3.4 from the Gamow-Teller
operator and 1.2 from the Fermi operator for SkO$'$, and 3.7 from the
Gamow-Teller operator and 1.0 from the Fermi operator for SkM*.  

In summary, the ease with which the GCM works in a large model space, even with
several coordinates, means that it can include physics that is beyond the shell
model.  And because it mixes mean fields and has no issues with phase
transitions, it offers a more comprehensive and accurate treatment of
correlations than does the QRPA.  One direction for future research in the
\textit{pn} GCM is a more complete effective interaction in multi-shell model
spaces; another, perhaps more important, is an implementation together with
Skyrme and Gogny energy-density functionals or with their successors.  The
combination of projection and GCM with density-functional theory poses a few
conceptual problems (see, e.g., Ref.\ \cite{ben11}) but recent progress
suggests that they will be resolved before too long.  The inclusion of
\textit{pn} degrees of freedom as generator coordinates should soon improve the
quality of density-functional-based double-beta matrix elements.

We gratefully acknowledge useful discussions with T. R. Rodr\'iguez.  This
work was supported by the U.S.\ Department of Energy through Contract No.\
DE-FG02-97ER41019, and JUSTIPEN (Japan-U.S. Theory Institute for Physics with
Exotic Nuclei) under Grant No. DE-FG02-06ER41407 (U. Tennessee).  We used
computational resources at the National Institute for Computational Sciences
(http://www.nics.tennessee.edu/).

\end{document}